\documentclass{ceurart}

\usepackage{graphicx} 
\usepackage{subcaption}
\usepackage{float}
\usepackage[frozencache=true, cachedir=minted-output]{minted} 
\begin{document}

\copyrightyear{2021}
\copyrightclause{Copyright for this paper by its authors.
  Use permitted under Creative Commons License Attribution 4.0
  International (CC BY 4.0).}

\conference{DLCP'21: International Workshop on Deep Learning in Computational Physics, June 28--29, 2021, Moscow, Russia}

\title{Application of Deep Learning Technique to an Analysis of Hard Scattering Processes at Colliders}

\author[1]{Lev Dudko}[
email=lev.dudko@cern.ch
]
\author[1]{Petr Volkov}[
email=petr.volkov@cern.ch
]
\author[1]{Georgii Vorotnikov}[
email=georgii.vorotnikov@cern.ch
]
\author[1]{Andrei Zaborenko}[
email=andrei.zaborenko@cern.ch
]
\address[1]{Skobeltsyn Institute of Nuclear Physics, M.V. Lomonosov Moscow State University,
  1(2) Leninskie gory, Moscow 119991, Russian Federation}

\begin{abstract}
 Deep neural networks have rightfully won the place of one of the most accurate analysis tools in high energy physics.  In this paper we will cover several methods of improving the performance of a deep neural network in a classification task in an instance of top quark analysis. The approaches and recommendations will cover hyperparameter tuning, boosting on errors and AutoML algorithms applied to collider physics.
\end{abstract}


\maketitle

\section{Physics task}

In the scope of high energy physics analysis the measurement of t-channel single top-quark production is used as a benchmark to calibrate the analytical tools and assumptions. Then the developed methods are used to measure deviations from the Standard Model -- the Flavor Changing Neutral Currents (FCNC). The physics task is similar to the analysis of CMS collaboration~\cite{CMS:2016uzc}.
Neural networks are used extensively throughout the analysis to separate background and signal events. First neural network model is used to filter out the multi-jet QCD events as these events are hard to model with existing Monte-Carlo methods. This network uses only five variables as its inputs and has relatively small number of trainable parameters. After the multi-jet QCD suppression a larger Standard Model neural network is used to identify top-quark events. These two different tasks allow us to test the performance of described methods in two separate instances.


\section{DNN hyperparameter tuning}
\subsection{Overview}

Most of machine learning models have two types of parameters: trainable and non-trainable\cite{omalley2019kerastuner}. Trainable parameters change during training, and non-trainable (or hyperparameters) are set by the user. For example, in a typical multi-layer perceptron hyperparameters can be the number of hidden layers, number of neurons in each hidden layer, the learning rate, regularization constants, etc. In this case trainable parameters are the weights and biases of each neuron.

Different hyperparameter combinations give the deep learning model varying degree of complexity and non-linearity. Therefore, hyperparameter tuning can help resolve overfitting and underfitting to a certain degree. 

Finding the best combination of hyperparameters can be a challenging task as the model's performance can only be evaluated when the training is finished. Luckily, many hyperparameter optimization frameworks exist to automate this tedious process.

\subsection{Hyperparameter tuning frameworks}

One of the most established and well-known frameworks is called \texttt{Optuna}\cite{optuna_2019}. It provides tools to tune any machine learning model and quickly visualize the results. Another useful feature of \texttt{Optuna} is the budget -- the amount of time used to tune a single model. This can be useful to compare different models and still give both fair treatment, tuning them for the same time. 

However, we have opted to use \texttt{Keras Tuner}\cite{omalley2019kerastuner} for its tight integration with \texttt{Tensorflow Keras}, allowing us to use hyperparameter tuning with minimal code modifications and dependencies. After each trial this module generates a \texttt{.JSON} file containing all required information about a single run. If the user wishes not only get the best performing hyperparameter combination but also to explore the dependencies and tendencies of their deep learning model, the results can be parsed and visualized. 

\subsection{Tuner setup}

The first major step in setting up any hyperparameter tuning is defining all possible hyperparameter combinations -- the hyperparameter space. Usually the ranges of numeric hyperparameters are defined either with a distribution or with a set array of values. The latter is done by defining the minimum and maximum values and setting the step parameter, the distance between two consecutive samples in the range. We have used this approach to visualize the relations between the model's performance and the values of its hyperparameters. The non-numerical hyperparameters (the hidden layers activation function, for example) are chosen with the \texttt{Choice} method. 

The second step is to define a score variable -- a metric to quantify and compare the model's performance. In the default case this variable can be equal to model's loss value (binary crossentropy in our binary classification task) or to any pre-made or user-defined metric. 

With the hyperparameter space expanding with each new variable, a suitable algorithm to navigate it is required. \texttt{Keras Tuner} provides three basic Tuner algorithms, each with its own advantages and drawbacks: \texttt{BayesianOptimization}, \texttt{Hyperband} and \texttt{RandomSearch}. 

\texttt{BayesianOptimization} algorithm uses tuning with Gaussian process. This is the fastest built-in algorithm, however, it can only find the local minima in the hyperparameter space. In our tests it converged within approximately 10\% of the total combinations in the hyperparameter space. The parameters this algorithm converged on were adequate albeit not the overall best.

\texttt{Hyperband} algorithm uses the performance of the first epochs to compare different hyperparameter combinations. We decided against this method as models with different learning rates will have different \texttt{Hyperband} performance which will not reflect their overall accuracy. 

\texttt{RandomSearch} tuning algorithm randomly samples hyperparameter combinations from the hyperparameter space. This method does not use any fancy logic, however, it reliably provides a near-best result when covering 40-50\% of the hyperparameter space. 

The code covering the needed adaptations is available in the Appendix.

\subsection{Results interpretation}

After the tuning process is compete, all trials results are stored in \texttt{.JSON} files. The user can opt to use the best configuration without looking at other models, however, plotting the relations can provide useful insights into how the chosen model is performing. 

\begin{figure}
\centering
\includegraphics[width=0.49\linewidth]{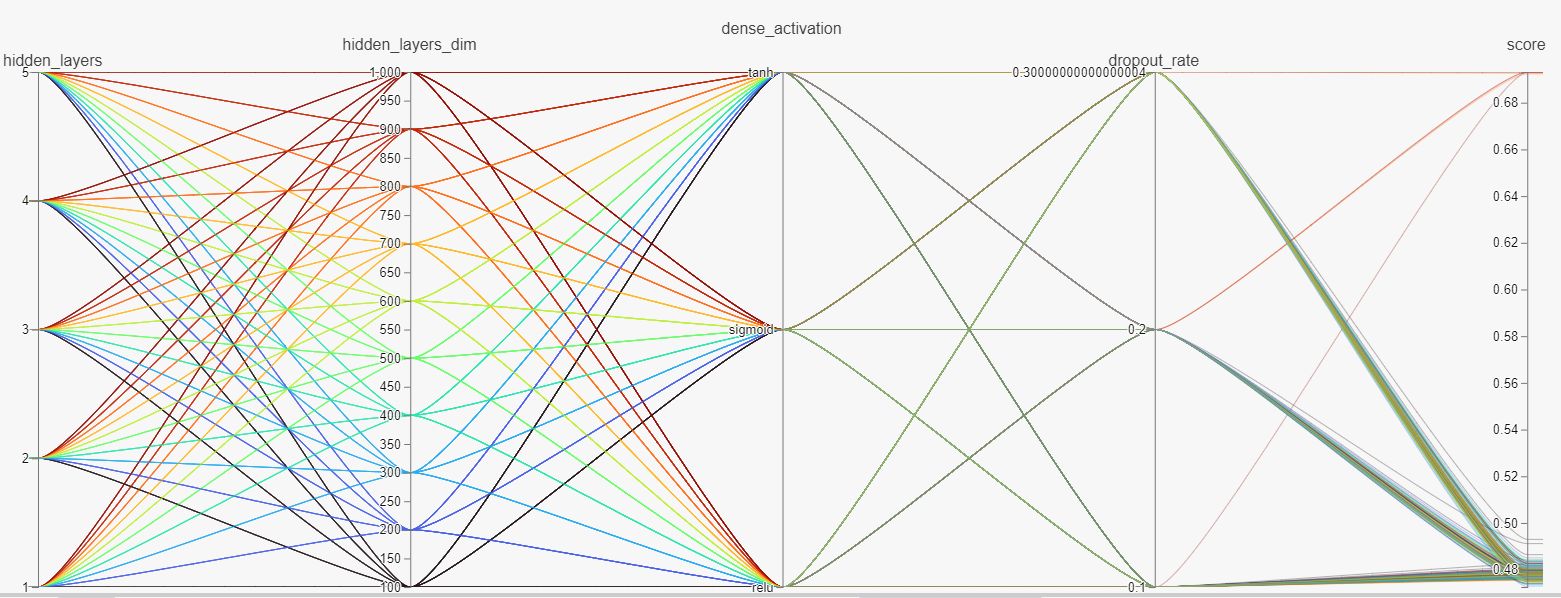}
\includegraphics[width=0.49\linewidth]{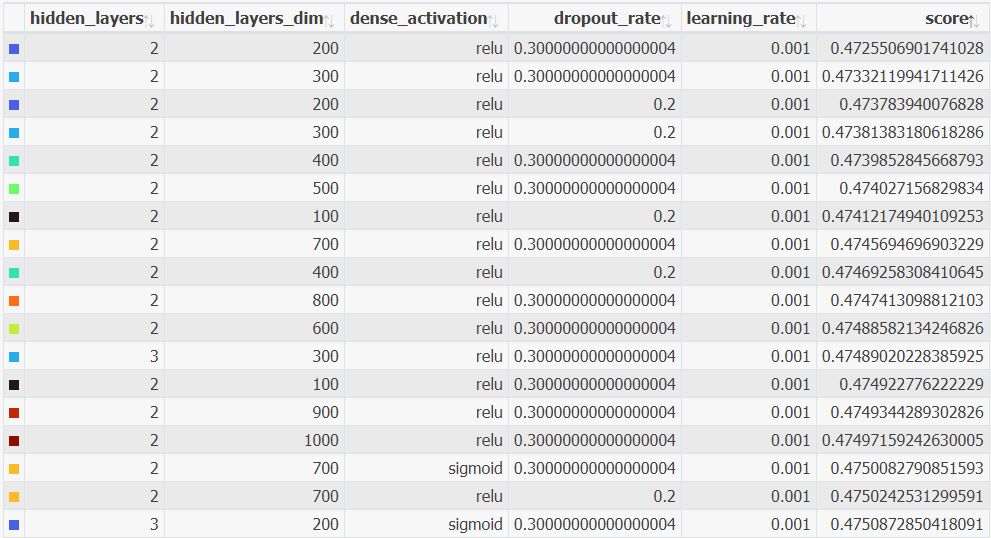}
\caption{Example of the hiplot interface}
\label{fig:hiplot_fig}
\end{figure}

For general overview one can use Facebook's \texttt{hiplot}\cite{hiplot} utility (Figure~\ref{fig:hiplot_fig}). This interface allows the user to quickly analyse the trained models, sort them by their performance and check specific hyperparameter combinations. 

\begin{figure}[h]
        \begin{subfigure}[b]{0.33\textwidth}
                \centering
                \includegraphics[width=\linewidth]{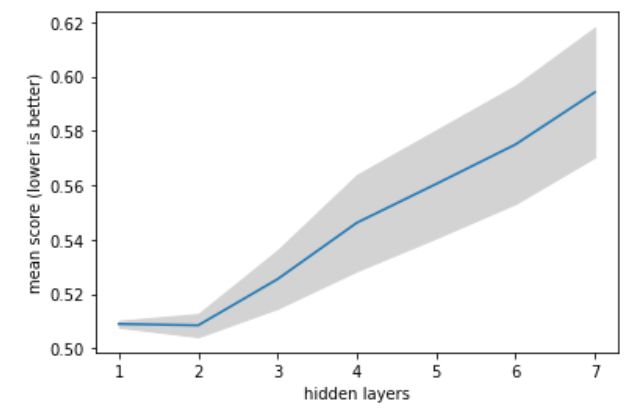}
        \end{subfigure}%
        \begin{subfigure}[b]{0.33\textwidth}
                \centering
                \includegraphics[width=\linewidth]{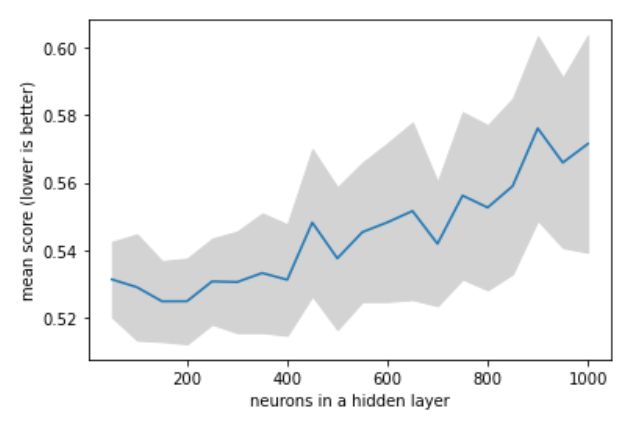}
        \end{subfigure}%
        \begin{subfigure}[b]{0.33\textwidth}
                \centering
                \includegraphics[width=\linewidth]{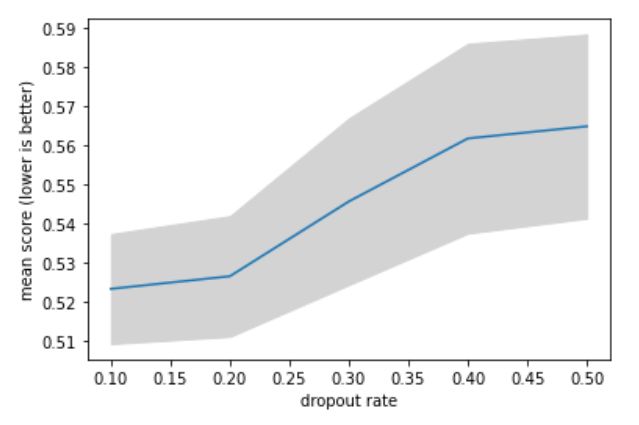}
        \end{subfigure}%
        \caption{One-dimensional plots describing the relations between hyperparameter's values and model's performance for the Standard Model neural network}
\label{fig:tuner_plot_SM}
\end{figure}

\begin{figure}[h]
        \begin{subfigure}[b]{0.33\textwidth}
                \centering
                \includegraphics[width=\linewidth]{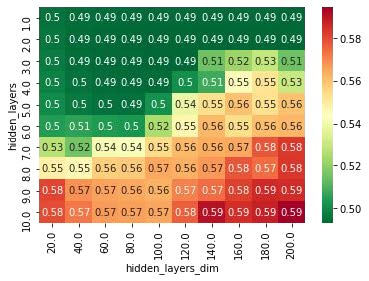}
        \end{subfigure}%
        \begin{subfigure}[b]{0.30\textwidth}
                \centering
                \includegraphics[width=\linewidth]{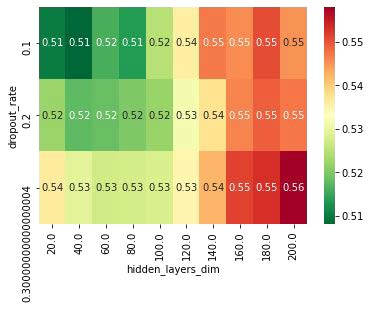}
        \end{subfigure}%
        \begin{subfigure}[b]{0.35\textwidth}
                \centering
                \includegraphics[width=\linewidth]{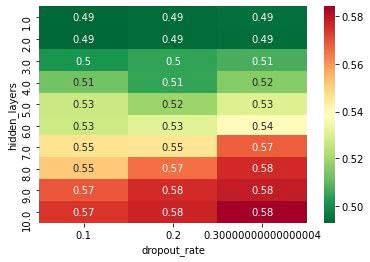}
        \end{subfigure}%
        \caption{Two-dimensional plots describing the relations between hyperparameter's values and model's performance for the QCD suppression neural network}
\label{fig:tuner_plot_QCD}
\end{figure}

To further investigate the hyperparameter space, one can plot the relations between models' performance and the values of used hyperparameters. We give two examples of such visualization in Figures~\ref{fig:tuner_plot_SM} and  ~\ref{fig:tuner_plot_QCD}. In the first set of plots covering the tuning of larger Standard Model neural network we demonstrate the relation between the model's performance and a certain hyperparameter value, averaging over the rest of hyperparameters. In the second set we used heatmaps to describe hyperparameter combinations for the smaller QCD suppression neural network.

Having investigated the hyperparameter space for two typical High Energy Physics tasks, we can give broader recommendations for neural network design for this field. First of all, using \texttt{ReLU} for hidden layers activation function is advisable. Standard \texttt{tanh} and \texttt{sigmoid} functions lead to worse performance in deeper, bigger networks. In both cases networks with one or two hidden layers performed better and more stable than their deeper counterparts. The number of nodes in the hidden layers varied depending on the amount of input features: for the bigger network with 50 input features the amount of neurons lied in range between 200 and 400, and for the smaller network with 5 input features it was closer to 120.

\section{AutoML}
\subsection{Overview}

AutoML approach covers finding the optimal machine learning model, training it, evaluating it and tuning it if its performance is insufficient. In theory, given enough time and computational resources, this approach can yield an adequate model without investing researcher's time into complex architecture tuning and feature engineering. 
High energy physics data is close in structure to tabular data. In other areas tabular data may contain text and categorical data, but in high energy physics data is primarily numerical and can be organized into columns, so the task is simpler in a certain way.

\begin{figure}[!th] 
\begin{subfigure}{0.48\textwidth}
\includegraphics[width=\linewidth]{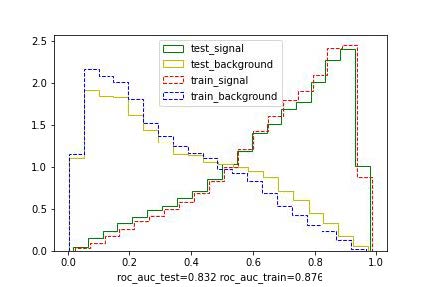}
\caption{Discriminator of the \texttt{Explain} mode model} \label{fig:a}
\end{subfigure}\hspace*{\fill}
\begin{subfigure}{0.48\textwidth}
\includegraphics[width=\linewidth]{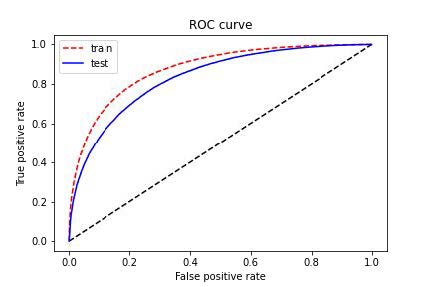}
\caption{ROC Curve of the \texttt{Explain} mode model} \label{fig:b}
\end{subfigure}

\begin{subfigure}{0.48\textwidth}
\includegraphics[width=\linewidth]{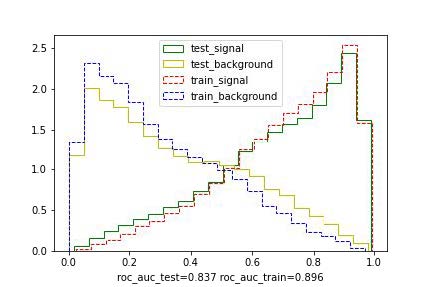}
\caption{Discriminator of the \texttt{Compete} mode model} \label{fig:c}
\end{subfigure}\hspace*{\fill}
\begin{subfigure}{0.48\textwidth}
\includegraphics[width=\linewidth]{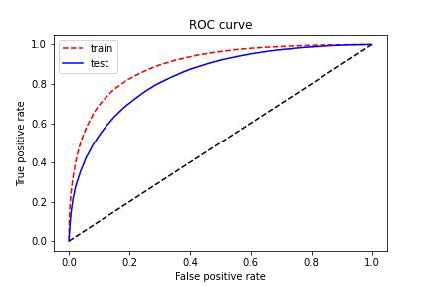}
\caption{ROC Curve of the \texttt{Compete} mode model} \label{fig:d}
\end{subfigure}

\begin{subfigure}{0.48\textwidth}
\includegraphics[width=\linewidth]{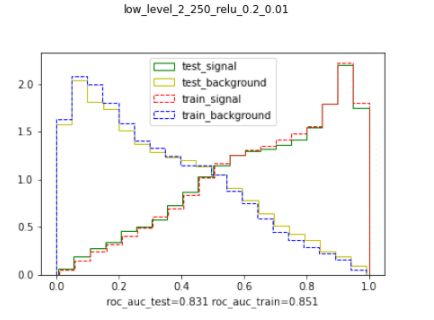}
\caption{Discriminator of the tuned Tensorflow Deep Neural network} \label{fig:e}
\end{subfigure}\hspace*{\fill}
\begin{subfigure}{0.48\textwidth}
\includegraphics[width=\linewidth]{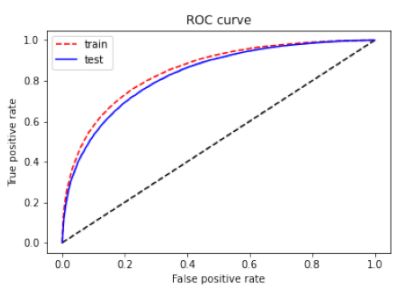}
\caption{ROC Curve of the tuned Tensorflow Deep Neural network} \label{fig:f}
\end{subfigure}

\caption{Comparison between two AutoML models and a tuned DNN model} \label{fig:1}
\end{figure}

\subsection{Deep learning for tabular data}
Deep learning has shown comparable performance\cite{gorishniy2021revisiting} in tabular data classification to gradient boosting models (\texttt{CatBoost, LightGBM}), which are much cheaper in terms of computational resources. However, our preliminary testing done using the \texttt{LightGBM} package showed that fine-tuned neural network performs slightly better and overfits less than a tuned \texttt{LightGBM}\cite{lgbm} model.
Google's \texttt{Adanet}\cite{cortes2017adanet} package uses genetic algorithms to create a custom neural network structure for each machine learning task. This approach has a promising idea, however, the lack of support for weighted events limits its uses in high energy physics analysis where every event has a very specific weight value. 

\subsection{AutoML in high energy physics}
We have opted for using \texttt{mljar-supervised}\cite{mljar} library for automated machine learning. This library is easy to comprehend, has several performance modes (focused on data exploration, speed of inference or maximum accuracy of classification).
It uses several machine learning algorithms (Linear, Random Forest, Extra Trees, LightGBM, Xgboost, CatBoost, Neural Networks, and Nearest Neighbors) for classification and then creates an ensemble of best performing models for final classification.
Here we present the results of two AutoML models in \texttt{Explain} (runs in a dozens of minutes and performs Exploratory Data Analysis) and \texttt{Compete} (maximum classification accuracy, needs more computational time, we have run it for a day) modes as the speed of inference is not crucial in the current analysis.
The performance comparisons are shown in Figure~\ref{fig:1}. Both classification modes provided good accuracy and even outperformed the tuned neural network in terms of \texttt{roc\_auc} metric on the test dataset. However, this was done with a much higher degree of overfitting, thus reducing the AutoML model's predictive power.

\subsection{Resume}
\texttt{mljar-supervised} provided a good baseline in all our use-cases, with its maximum accuracy mode overfitting a bit more that we would like it to. As it is much easier to control overfitting inside Tensorflow package through regularization and early stopping callback, for the time being we will continue to use it in our analysis, but this AutoML package came close to it in terms of classification accuracy.  We will definitely monitor the development of this great tool and continue testing it.

\section{DNN boosting on errors}

We have also tried boosting on errors. The concept of this method is simple:

\begin{enumerate}
  \item Train a machine learning model on unaltered data, get its predictions
  \item Take the events that were miss-labelled after first classification and artificially increase their weights
  \item Train a second model using the 'corrected' weight vector, supposedly achieving better performance, as this model will put more emphasis on difficult events that were hard to differentiate in the first place
  \item Update the weights and repeat
\end{enumerate}

\begin{figure}[h]
        \begin{subfigure}[b]{0.25\textwidth}
                \centering
                \includegraphics[width=\linewidth]{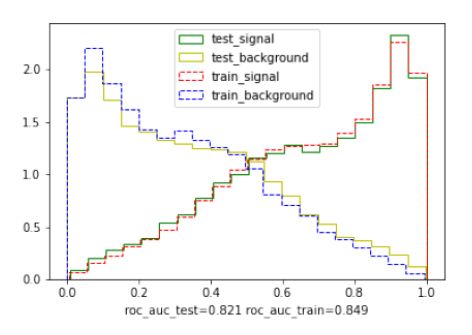}
                \caption{Baseline}
        \end{subfigure}%
        \begin{subfigure}[b]{0.25\textwidth}
                \centering
                \includegraphics[width=\linewidth]{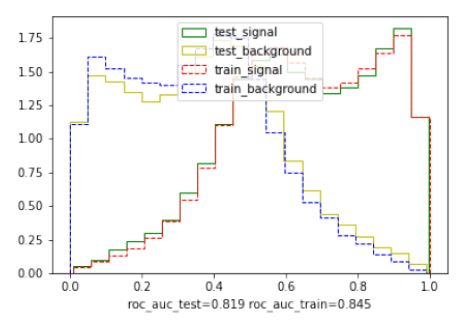}
                \caption{First iteration}
        \end{subfigure}%
        \begin{subfigure}[b]{0.25\textwidth}
                \centering
                \includegraphics[width=\linewidth]{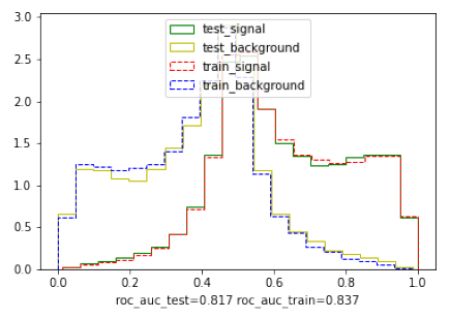}
                \caption{Second iteration}
        \end{subfigure}%
        \begin{subfigure}[b]{0.25\textwidth}
                \centering
                \includegraphics[width=\linewidth]{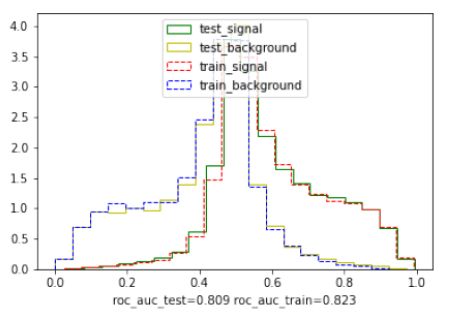}
                \caption{Third iteration}
        \end{subfigure}
        \caption{Performance degradation with each boosting on errors iteration}
\label{fig:2}
\end{figure}

This method did not work as the results of classification worsened after each iteration. The illustration of this performance degradation can be found on Figure~\ref{fig:2}. The best explanation we have come up with was that deep neural network is not a weak learner, which were noticed to benefit from such manipulations.

\section{L-regularisation}

Through experimentation with input features we have found out that certain features worked extremely well for first-order classification, causing the model to increase its weights associated with these features, in turn lowering the importance of other, still needed, features. The proposed method included using l-regularization~\cite{l2_reg} to limit high weights so that they will not overshadow other weights as much. 

We have conducted the l-regularization study by getting a good baseline model from the \texttt{Keras Tuner}, fixing its hyperparametes and varying only the regularization constant. This was done for L1, L2 and L12 regularization types.

\begin{figure}[h]
        \begin{subfigure}[b]{0.33\textwidth}
                \centering
                \includegraphics[width=\linewidth]{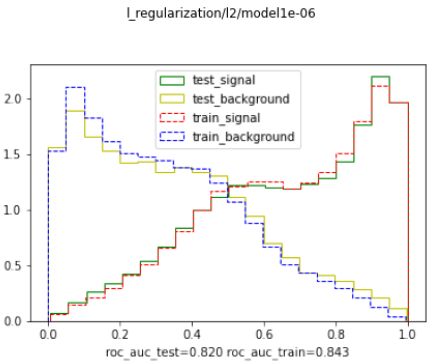}
                \caption{The constant is too small}
        \end{subfigure}%
        \begin{subfigure}[b]{0.33\textwidth}
                \centering
                \includegraphics[width=\linewidth]{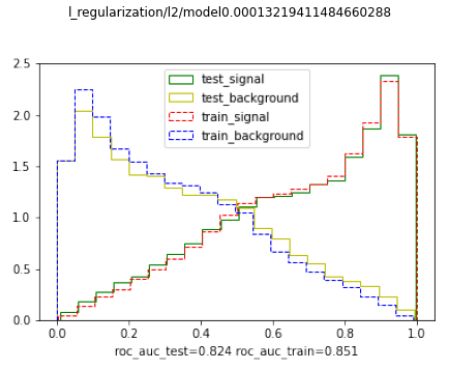}
                \caption{The constant is chosen well}
        \end{subfigure}%
        \begin{subfigure}[b]{0.38\textwidth}
                \centering
                \includegraphics[width=\linewidth]{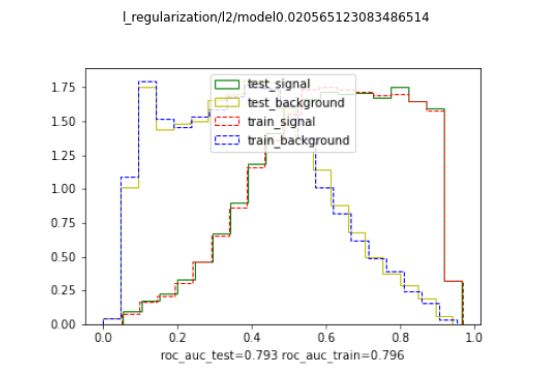}
                \caption{The constant is too large}
        \end{subfigure}%
        \caption{DNN model's dicriminators. The only difference between the models is the regularization constant.}
\label{fig:3}
\end{figure}

The results showed that when the regularization constant is chosen right, the discriminant distribution curve becomes smoother and general classification performance increases. However,  when the regularization constant is too small, there are no perceivable improvements. High regularization constant values can be even more detrimental as they will outright decrease the classification performance of otherwise decent model. The illustration of these relations can be found in Figure~\ref{fig:3}.

We have not noticed considerable differences between the regularization types, L2 performed slightly better, but that disparity was within the margin of error.

\section{Conclusion}
We have demonstrated several approaches to improve the accuracy of classification based on a model of a Deep Neural Network in High Energy Physics. Here we provide a short summary of the methods described in the paper. DNN hyperparameter tuning is an effective method of improving the accuracy of the model, but it requires a lot of computational resources. The required computing time can be reduced by making hyperparameter space smaller and using a suitable optimization algorithm. We also provided recommendations based on our hyperparameter space exploration for typical high energy physics datasets. AutoML for tabular data can be used in High Energy Physics with relative success and little machine learning experience, however, the degree of overfitting is hard to control. Boosting on errors is not advised to use with DNNs, and using any type of L-regularization is advisable if the regularization constant is chosen correctly.

\bibliography{sample-ceur}
\newpage
\appendix

\section{Code for Keras Tuner adaptation}

The function that returns the base model: 

\begin{minted}[breaklines, frame=single]{python}
def createModel(dim, hidden_layers, hidden_layers_dim, dense_activation, dropout_rate, learning_rate):
    model = Sequential() model.add(Input(shape=(dim,))) 
    for n in range (hidden_layers):
        model.add(Dense(hidden_layers_dim, activation=dense_activation))
        model.add(Dropout(rate = dropout_rate))
    model.add(Dense(units=1, activation='sigmoid'))
    adam = Adam(lr=learning_rate)
    model.compile(loss='binary_crossentropy', optimizer=adam, metrics=['mean_squared_error'])
    return model
\end{minted}

The function that returns the model adapted to the Keras Tuner environment:

\begin{minted}[breaklines, frame=single]{python}
def build_model(hp):
    model = Sequential()
    model.add(Input(shape=(dim,)))
    for n in range (hp.Int('hidden_layers', min_value = 1, max_value = 7, step = 1)):
        model.add(Dense(units = hp.Int('hidden_layers_dim', min_value = 50, max_value = 1000, step = 50), activation = hp.Choice('dense_activation',values=['relu', 'tanh'])))
        model.add(Dropout(rate = hp.Float('dropout_rate', min_value = 0.1, max_value = 0.5, step = 0.1)))
    model.add(Dense(units=1, activation='sigmoid'))
    adam = Adam(hp.Choice('learning_rate', values=[1e-2, 1e-3])))
    model.compile(optimizer=adam, loss='binary_crossentropy')
    return model
\end{minted}

\end{document}